\documentclass[]{spie}  

\usepackage{amsmath,amsfonts,amssymb}
\usepackage{graphicx}
\usepackage[colorlinks=true, allcolors=blue]{hyperref}
\usepackage{nccmath}
\usepackage{siunitx}
\usepackage{float}

\title{Testing of the LSST's photometric calibration strategy at the CTIO 0.9 meter telescope}

\author[a]{Michael W. Coughlin}
\author[b]{Susana Deustua}
\author[c]{Augustin Guyonnet}
\author[c,d]{Nicholas Mondrik}
\author[e]{Joseph P. Rice}
\author[c,f]{Christopher W. Stubbs}
\author[e]{John T. Woodward}

\affil[a]{Division of Physics, Math, and Astronomy, California Institute of Technology, Pasadena, CA 91125, USA}
\affil[b]{Space Telescope Science Institute, 3700 San Martin Drive,
Baltimore, MD 21218, USA}
\affil[c]{Department of Physics, Harvard University, Cambridge, MA 02138, USA}
\affil[d]{LSSTC Data Science Fellow}
\affil[e]{National Institute of Standards and Technology (NIST),
100 Bureau Drive, Gaithersburg MD 20899}
\affil[f]{Department of Astronomy, Harvard University, Cambridge MA 02138, USA}

\authorinfo{Further author information: (Send correspondence to Michael W. Coughlin) E-mail: mcoughli@caltech.edu}

\begin{document} 
\maketitle

\begin{abstract}
The calibration hardware system of the Large Synoptic Survey Telescope (LSST) is designed to measure two quantities: a telescope's instrumental response and atmospheric transmission, both as a function of wavelength. First of all, a ``collimated beam projector'' is designed to measure the instrumental response function by projecting monochromatic light through a mask and a collimating optic onto the telescope. During the measurement, the light level is monitored with a NIST-traceable photodiode. This method does not suffer from stray light effects or the reflections (known as ghosting) present when using a flat-field screen illumination, which has a systematic source of uncertainty from uncontrolled reflections. It allows for an independent measurement of the throughput of the telescope's optical train as well as each filter's transmission as a function of position on the primary mirror. Second, CALSPEC stars can be used as calibrated light sources to illuminate the atmosphere and measure its transmission. To measure the atmosphere's transfer function, we use the telescope's imager with a Ronchi grating in place of a filter to configure it as a low resolution slitless spectrograph. In this paper, we describe this calibration strategy, focusing on results from a prototype system at the Cerro Tololo Inter-American Observatory (CTIO) 
0.9 meter telescope. We compare the instrumental throughput measurements to nominal values measured using a laboratory spectrophotometer, and we describe measurements of the atmosphere made via CALSPEC standard stars during the same run.
\end{abstract}


\section{Introduction}
\label{sec:introduction}

Precision, multiband photometry remains essential for many astrophysical endeavors, which include studying the expansion history of the universe with type Ia supernovae, using photometric redshifts to determine redshifts to galaxies and clusters, and detecting and characterizing exoplanets with transits \cite{BoGo2014}.
Photometric calibration consists of the observation of standard sources and the measurement of extinction coefficients in a program's passbands \cite{StDo2010,AlBu2016,Boh2016,NaAx2016,KrSu2017}. In addition, for ground-based observations, it also consists of correcting for instrumental, telluric and astrophysical effects.  
For precision calibration, a strong argument is made for breaking up the measurement into an in-dome measurement of the throughput of the optical system and a direct measurement of the optical transmission of the atmosphere \cite{StTo2006}.
For an imaging system, the instrument is a combination of the telescope optics and the camera, common to all observations, plus the filters.
In this paper, we describe progress on determining atmospheric and instrumental throughput measurements to improve precision photometry.

\begin{figure*}[t]
\hspace*{-0.5cm}
\centering
 \includegraphics[width=5in]{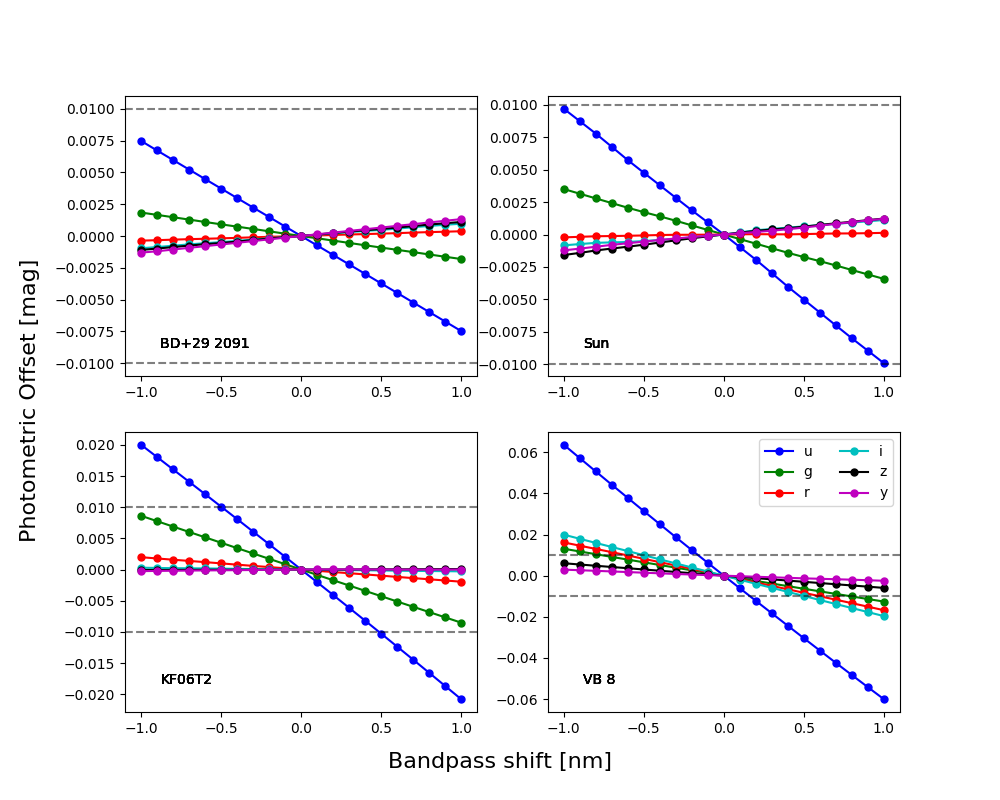}
 \caption{Photometric offsets induced by translations of the LSST filter bandpasses.  In this figure, "mag" stands for astronomical magnitude: dashed horizontal lines denote a 1\,\% offset.  Due to the shape of the stellar spectra, the \textit{u} filter is typically the most affected.  Spectral energy distributions (SEDs) are taken from the CALSPEC database\protect\footnotemark.} 
 \label{fig:bandpassshift}
\end{figure*}

Photometric calibration impacts many scientific applications. Error in calibration can result in systematic variations in photometric colors of galaxies, photometric redshifts, outlier fractions (percentage of supernovae with photometric redshifts showing significant deviations from their measured spectroscopic redshifts) and sample completeness. This error may be a function of altitude and azimuth, wavelength, seeing, and other time-dependent quantities, and is affected by observing strategy as well as uncertainties in atmospheric and galactic modeling, including the line of sight reddening and extinction due to the interstellar medium.
As an example of how a shift in the effective bandpass of an optical filter in an otherwise ideal system can impact photometric measurements, figure~\ref{fig:bandpassshift} shows the induced offset in magnitudes on sample F, G, K, and M stars taken from the CALSPEC database \cite{BoDi2001,BoGo2014,BoMe2017}.
These systematics, if left untreated, represent a significant fraction of the error budget for next-generation astronomical surveys.
The optical transmission spectrum of the atmosphere and its impact on high-precision photometry has been previously described\cite{StHi2007}.
In this analysis, we present strategies to address the uncertainties contributed by the telescope and Earth's atmosphere.
\footnotetext{http://www.stsci.edu/hst/observatory/crds/calspec.html}

The goal of imaging is to determine the location and the flux of astronomical sources.
The detected flux in a given pixel $i$ is given as \cite{StTo2012}
\begin{equation}
S_i = \sum_{\mathrm{source} j} \int \phi_j(\lambda) R_i(\lambda) T(\lambda) G(\lambda) A_i d\lambda.
\end{equation}
where $\lambda$ is the wavelength of the light and $i$ is the pixel index,
$\phi_j(\lambda)$ is the spectral photon distribution of source $j$, $R_i(\lambda)$ is the pixel dimensionless system transmission function, which includes both the telescope and quantum efficiency (QE) of the pixel, $T(\lambda)$ is the transmission of the atmosphere, $G(\lambda)$ accounts for non-atmospheric extinction processes, and $A_i$ is the effective aperture of the pixel.
The goal of a calibration system is to measure $R_i(\lambda)$, which includes the effects of the optics, filters, and detectors and $T(\lambda)$, which measures the atmosphere.
Fringing, non-uniform interference filters, and water spots are some of the effects that change $R_i(\lambda)$ as a function of pixel, while Rayleigh scattering in the atmosphere, small particle scattering from aerosols, and molecular absorption drive differences in $T(\lambda)$ \cite{Houghton1977,Slater1980}. 

As a testing ground for our method, we use the Small and Moderate Aperture Research Telescope System (SMARTS) 0.9\,m telescope at CTIO.
This analysis is based on data obtained during a run from September-October 2017.
Observations from this run fit into two categories. First, in section~\ref{sec:CBPMeasurements}, we use a collimated beam projector to measure the throughput function of the SMARTS 0.9\,m telescope \cite{CoAb2016}. Second, in section~\ref{sec:atmosphere}, we measure the atmospheric transmission above the telescope by using a Ronchi grating to acquire low-resolution, slitless spectroscopy of standard stars between 400\,nm and 1\,$\mu$m.

\section{Collimated Beam Projector Measurements}
\label{sec:CBPMeasurements}

For the collimated beam projector measurement, we illuminate telescopes using a tunable, monochromatic light source whose flux is monitored using a calibrated NIST photodiode, which creates a fundamental metrology reference in order to determine the relative throughput of the telescope of interest.
This NIST photodiode is the standard to which we compare our measurement and is accurate to 0.1\,\% across the measurements here. 
The photodiode's uncertainty contributes via the tying of the photons emitted by the collimated beam projector system to the photodiode, which is especially important near 400\,nm and 1\,$\mu$m due to the limited system throughput.
The wavelength range of 400\,nm to 1\,$\mu$m allows us to securely cover many standard, photometric filters, which we will describe below.
This illumination system is used to project a grid of monochromatic stars of known brightness onto the focal plane, on which aperture or point spread function (PSF) photometry can be performed.
By comparing the amount of light detected by the charge-coupled device (CCD) to that projected from the system, we determine the wavelength dependent throughput of the telescope optical system, which includes: mirror reflectivity, filter and corrector optics transmission, and detector QE.
For a full description of the collimated beam projector, we refer the interested reader to our previous manuscript \cite{CoAb2016}.

The Sloan Digital Sky Survey used a similar system to characterize their camera only \cite{DoTa2010}.
A previous measurement at the Panoramic Survey Telescope and Rapid Response System (Pan-STARRS) telescope along these lines was performed \cite{StDo2010}.
There are a number of improvements made to previous measurements.
The previous measurement at the Pan-STARRS telescope had a shortcoming of creating a large (1$^\circ$) aperture illumination of the detector, which is incapable of differentiating ghosts in the optical system.
In that measurement, the Pan-STARRS optical system had a prominent optical ghost created from light bouncing off the CCD, off the back surface of the first corrector lens, and then refocused onto the CCD \cite{ToSt2012}.
This new system improves upon that system using multiplexed pinhole arrays.

\subsection{Filter Curve Extraction}
\label{subsec:filterCurveExtraction}
\begin{figure*}[th]
\centering
 \includegraphics[width=0.45\textwidth]{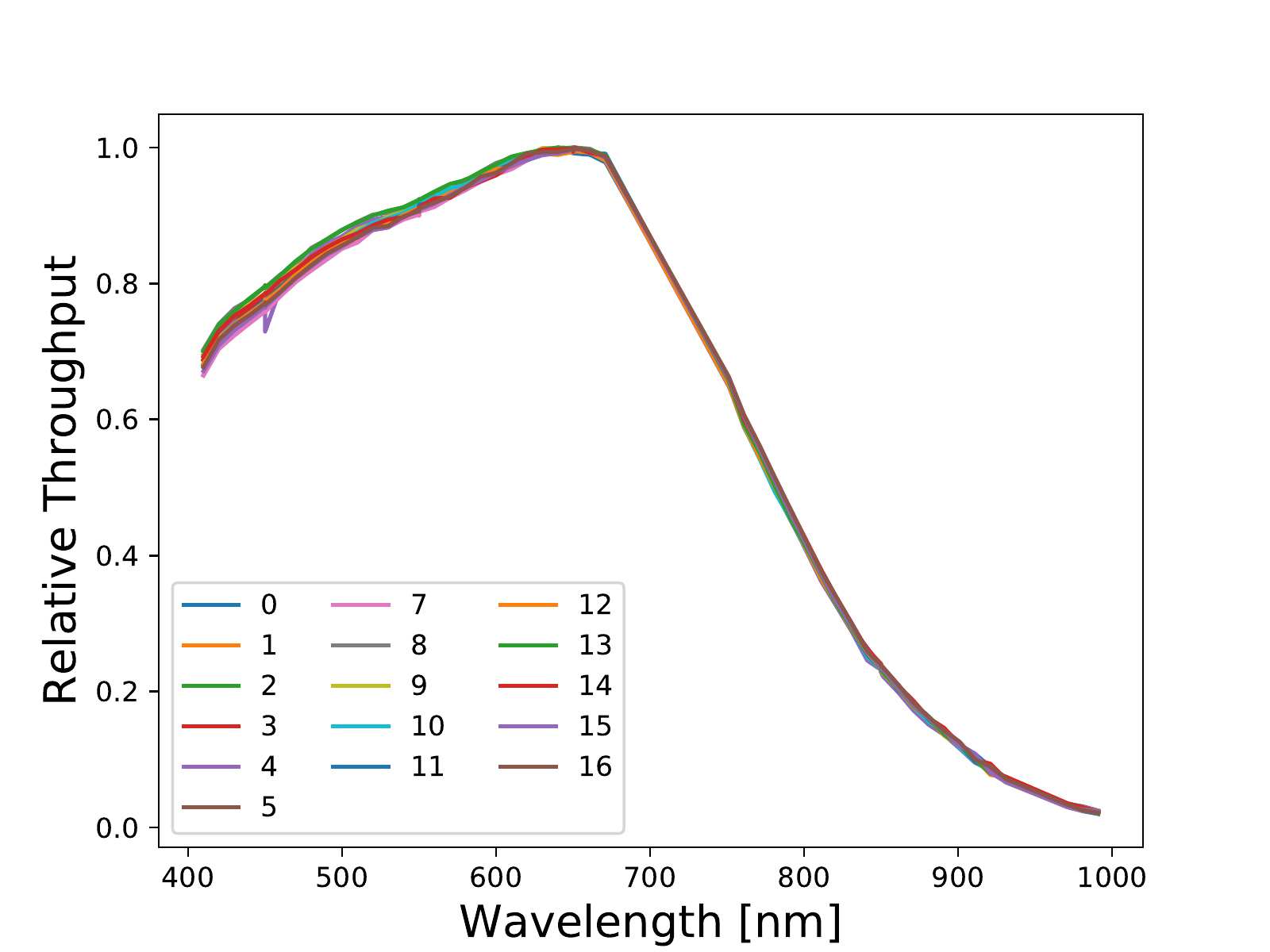}
 \includegraphics[width=0.45\textwidth]{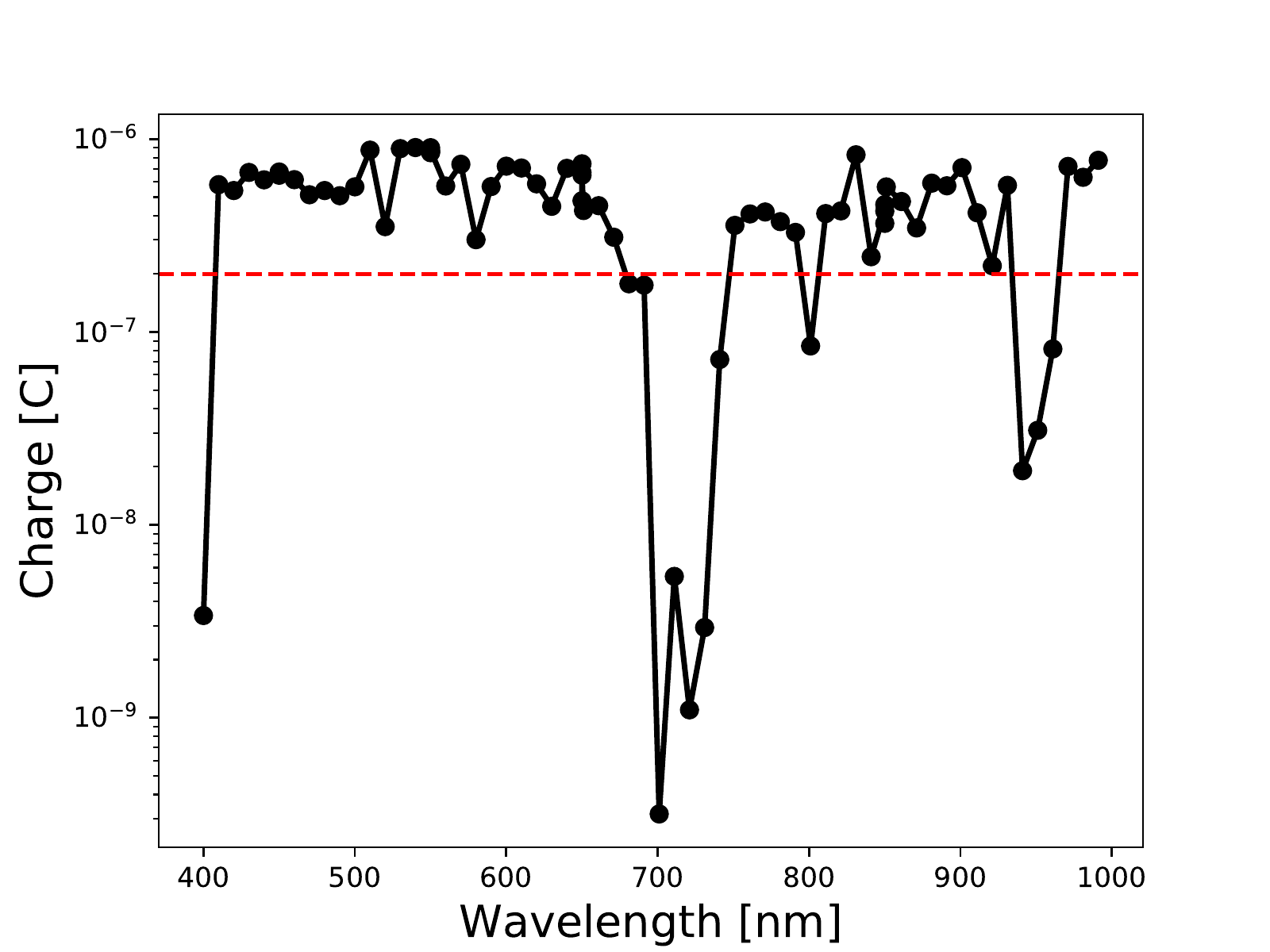} 
 \caption{On the left is the relative CTIO 0.9\,m telescope system throughput sample as a function of wavelength with no filter in the beam.  Each curve represents the measurement of a single spot on our projected image.  Points with injected charge levels on the monitor photodiode of less than $2  \times  10^{-7}$\,C are excluded.  Spot 6 is excluded due to contamination from defects on the detector. On the right is the injected charge as a function of wavelength for data shown on the left.  The laser dropouts are clearly shown as drops in accumulated charge.  The red line denotes the cutoff used to reject low signal-to-noise ratio (SNR) observations.}
 \label{fig:nofilthroughput}
\end{figure*}

We now determine the system throughput of the CTIO 0.9\,m telescope using the collimated beam projector. The imager is a 2048 by 2048 CCD camera from Scientific Imaging Technologies, Inc. (SITe) model SI424A.$^*$ Each pixel measures \SI{24}{\micro\metre} x \SI{24}{\micro\metre} and is delimited by channel stops in the serial direction and 3 gates in the parallel direction. The frame is split into 4 areas of 1024 by 1024, each one read by an amplifier located at each corner of the device. The field size is 13.6 arcminutes square, and the pixel size is 401 milliarcseconds/pixel. 

To determine the filter transmission, 
we first perform a measurement without filters in the beam, which provides a reference for the baseline transmission of the system.  The left of Figure~\ref{fig:nofilthroughput} shows the result of such a measurement. 
The first analysis presented is a broad system throughput scan with no filter in the beam. We scan the monochromatic line from an Ekspla NT242 Optical Parametric Oscillator (OPO) laser from 410\,nm to 990\,nm with 10\,nm steps.  We attempt to tune exposure times to produce a constant amount of charge on the monitor photodiode at each wavelength.  Due to the large variations in output power of the laser as a function of wavelength and the need to accomplish a scan within a reasonable amount of time (a few hours), we impose a maximum exposure time of 120 s, preventing us from achieving this charge level at all wavelengths.  The right of Figure~\ref{fig:nofilthroughput} shows the accumulated charge as a function of wavelength for data presented in the left of that figure. Additionally, due to issues with power dropouts in our tunable laser at specific wavelengths, we omit those points with a total integrated charge (as measured on the monitor photodiode) less than $2  \times  10^{-7}$\,C.  This cutoff is shown by the red dashed line. We also omit results for spot 6, as it is heavily contaminated by defects on the detector.  We also note that the spots are not necessarily placed in the same location on the focal plane in each scan.

\begin{figure*}[t]
\hspace*{-0.5cm}
\centering
 \includegraphics[width=6in]{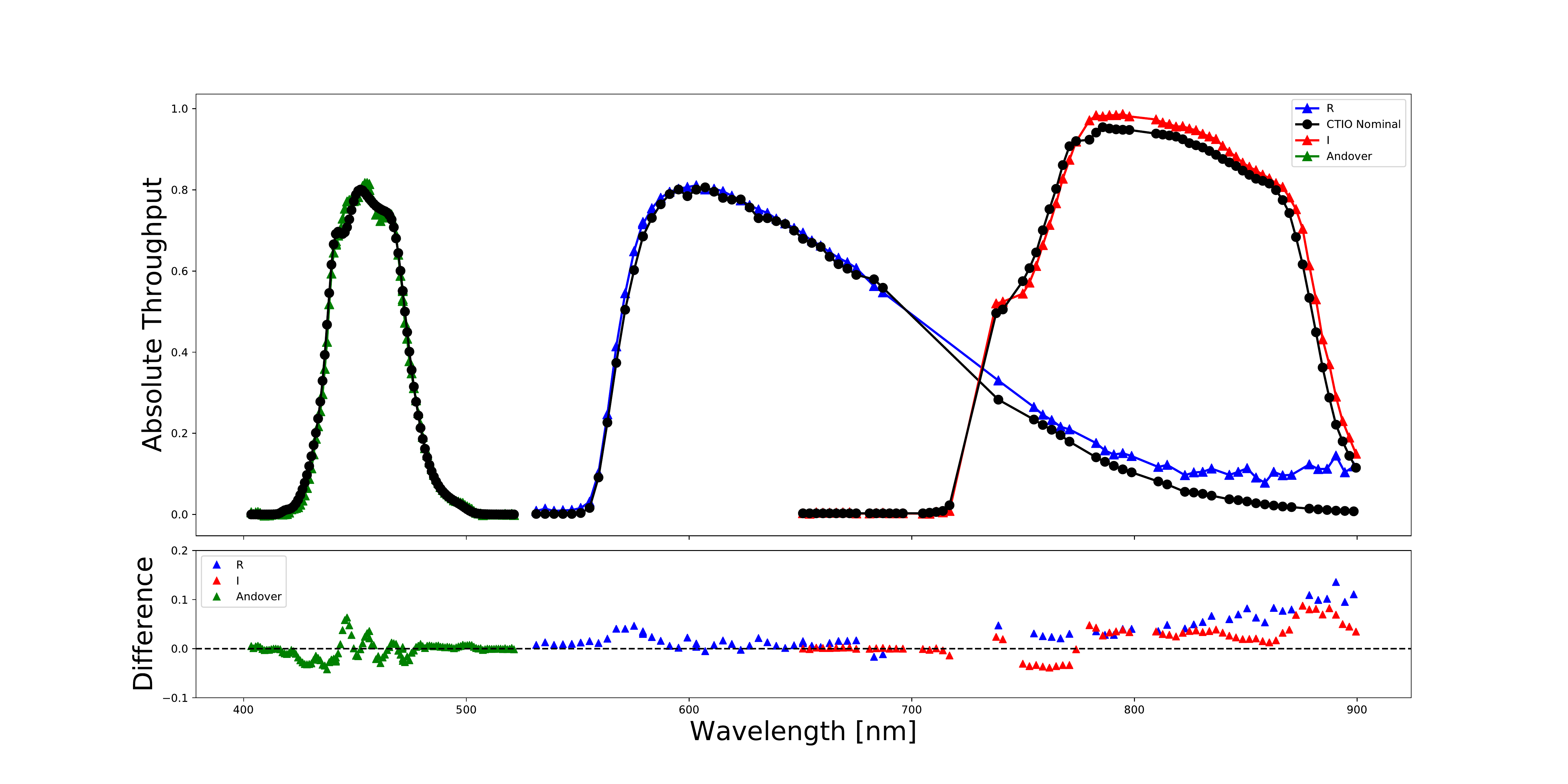}
 \caption{Top: Transmission curves for the Andover, R, and I filters.  Green (Andover), blue (R), and red (I) triangles show the measured throughputs for their respective filters, while the black circles denote the nominal values for transmission for each filter.  We are still working on deriving uncertainty estimates for our throughputs, but we expect a statistical standard uncertainty of at least $\pm$2\%. We expect that our throughputs are systematics-dominated at this time. Bottom: The difference between observed and nominal throughputs.  Note that points redward of 900\,nm have been omitted. See text for details.}
 \label{fig:throughputAll}
\end{figure*}

We then measure again with the filter in the beam; the ratio of the two provides an estimate of the filter bandpass at the measured wavelengths.
We perform these measurements of filter throughput for a number of filters: a custom ``\textit{g}'' filter\footnote{Andover model 450FSX40-50.}, as well as R and I.  The custom ``\textit{g}'' filter is needed, as the blue edge of the standard V filter is slightly too blue to scan effectively with our tunable laser light source. The second analysis presented in Figure~\ref{fig:throughputAll} shows the throughput curves of various filters at the telescope.  Measurements from our system are shown as green, blue, and red triangles. Black lines denote the nominal curve provided by Andover\footnote{\url{https://www.andovercorp.com/products/bandpass-filters/high-transmitting-bandpass-filters/}} for our ``\textit{g}'' filter, and the CTIO-provided nominal throughput curves for the R and I filters\footnote{\url{http://www.astro.gsu.edu/~thenry/SMARTS/0.9m.filters.pdf}}. 

\begin{figure*}[h]
\hspace*{-0.5cm}
\centering
 \includegraphics[width=5in]{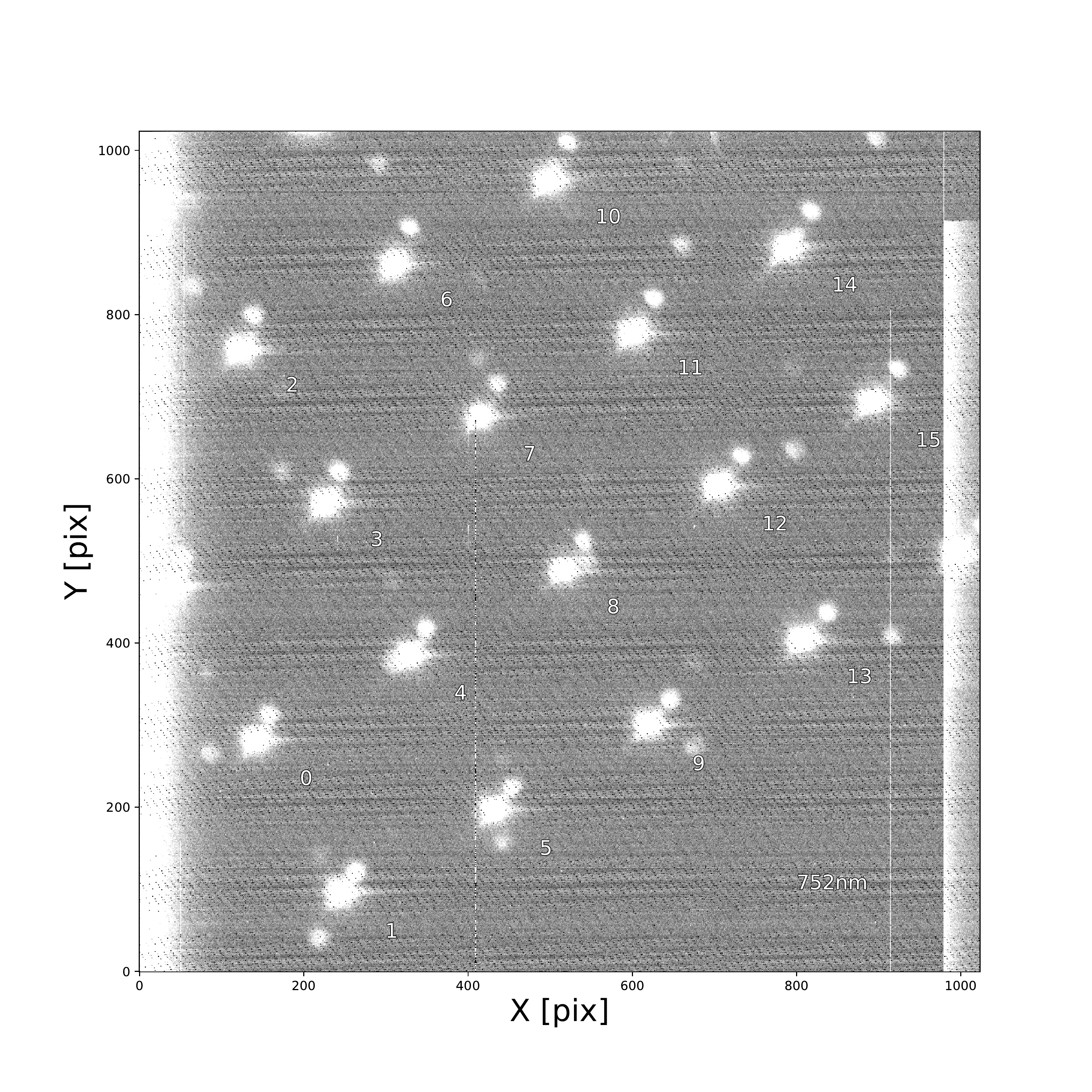}
 \caption{An example image from the I band scan presented in figure~\ref{fig:throughputAll}.  Ghosting due to internal reflections within the filter are readily visible next to the (labeled) main spots.  Without the use of a collimated beam, such ghosts would be unidentifiable.}
 \label{fig:CBPExample}
\end{figure*}

As described above, the collimated beam projector does not suffer from some of the same stray and scattered light effects (including contributions from ghosts) as non-collimated systems.
For example, any wavelength-dependent ghosting would masquerade as sensitivity variations in the system.  An example of this ghosting is shown in figure~\ref{fig:CBPExample}.
The resulting illumination correction can be a 5\,\% effect.
There are a number of potential sources of systematic error, however.
The first source of systematic error arises from ambient light in the dome. Any significant, unaccounted for spatial structure of the ambient light can bias our results, and masquerade as variations in the QE of the detector.  To remove this, we create a 2D background image, constructed out of 128 x 128 square pixel superpixels derived from medians of the science images. This is similar in purpose to a ``dark,'' but in this case the spatial structure of the ambient light can change quickly with time, such that using the science image itself is essential to perform the subtraction. This background image is subtracted prior to performing aperture photometry on the images.
In addition, errors in the wavelength calibration of the input source translate directly to errors in the measured throughput function. This is especially important near filter edges, where the transmission gradient with respect to wavelength can be extremely large.  To mitigate this, we pick off a small portion of the light exiting the system and monitor it with a spectrograph, which serves as a verification of the laser wavelength setting.

The monitor spectrograph also serves as an important check for light leaks into the system. The optical parametric oscillator in the tunable laser takes light at a single wavelength and creates two beams, known as ``signal'' and ``idler,'' where the output wave with higher frequency is the signal.
During our analysis, we discovered that approximately 1.5 \% of the signal beam from the tunable laser leaked into the output of the idler beam.
Our measurements at affected wavelengths (which appear to be those close to the degeneracy point of the laser at 710\,nm) are therefore compromised by this out-of-band light, and improvements to our system are necessary to mitigate this effect.  The affected wavelengths span approximately 713 nm to 743 nm.  Additionally, after our run, we discovered that during our scan of the I band, the tunable laser ceased to change wavelengths, becoming stuck at 900\,nm for the duration of the run.  Thus, any points taken after 900\,nm are taken with the correct exposure time, but with 900\,nm light being injected into the system, resulting in unrealistic measurements.  For this reason, we exclude all measurements taken after 900\,nm.  This includes points at 775\,nm, 800\,nm, and 850\,nm, which were taken at the end of the scan as a check on reproducibility.

\section{Determination of the atmospheric transparency}
\label{sec:atmosphere}

The strategy to determine optical atmospheric transmission is to exploit natural celestial sources with known Spectral Energy Distributions (SEDs) to backlight the atmosphere. Images of these sources are dispersed using a Ronchi grating installed in one of the two filter wheels of the SMARTS 0.9\,m telescope.  A Ronchi grating is used due to flatness of its dispersion efficiency with wavelength.


The image reduction process to extract spectra from the raw observations is described in the following subsection. The information is stored in the spectra in two ways: on a per-spectrum basis, and from comparison of spectra in a time series. While in both cases the SED of the targeted object needs to be known, the time series method alleviates requirements on telescope throughput calibration: it only relies on stable instrumental conditions, which are readily achievable over timescales of a night, but more difficult over the course of a several-year long survey, such as LSST.

\subsection{Atmospheric reductions}

An automated image reduction pipeline has been implemented to extract the spectra from the dispersed images. The first step of the image reduction is to subtract the median overscan and to trim the amplifiers region. The next step is to build a master-bias by median-stacking the bias images. The master-bias is subtracted to all the images so as to remove stable patterns introduced by the electronic readout.

The second step is to extract a rectangular box (the footprint) around the spectrum.  It is larger than the 2-D spectrum, and contains signal and  background (sky) pixels. The footprint is determined by combining the position of the 0th order image of the star and the direction of dispersion.
From a footprint, the spectrum profile is extracted using several methods: an aperture measurement or fitting a variety of functions, including a Gaussian, a Voigt and a Moffat. While the Voigt profile is a better approximation of the actual profile, the fitting is unstable. The standard deviation $\sigma$ of the Gaussian, although underestimating the width of the spectrum, produces stable estimates as a function of wavelength, and therefore was used in the following. It is used to estimate the variation of the PSF in the transverse direction as a function of the distance from the center of the field of view.

Hot pixels or cosmic rays in the footprint are flagged using the Laplacian detection method \cite{Dok2001}.
The background statistics, $\sigma$-rms and mean, are determined from the pixels in the footprint belonging to the sky map derived from source extractor (\emph{sextractor}). Pixel to pixel variation which exceed 5 $\sigma$-rms variation, or that is sharper than the seeing scale, are flagged, and optionally replaced by a local mean.

We now explore the idea of flatfielding this data set. Flatfielding  refers to any multiplicative task performed on the image. In the ideal case, a set of monochromatic and uniform illuminations can be used at once to correct both for pixel-to-pixel sensitivity variation - due to imperfections and defects on the telescope optical surface -  and for the overall QE variation of the detector as a function of wavelength and position. For dispersed images, different positions receive light from different wavelengths. The flatfielding task aims at recovering the illumination pattern from the dispersion of the first order of the grating by normalizing the instrumental response as a function of position and wavelength,
\begin{equation}
N(\vec{r}, \lambda) =  B(\vec{r}, \lambda) \cdot  T(\vec{r}, \lambda) \cdot QE(\vec{r}, \lambda)/g,
\end{equation}
where $B$ is the light intensity, $T$ is the transmission of the telescope, and $g$ is the gain of the camera. The monochromatic flatfields, with a known $B(\vec{r}, \lambda)$ are combined to build synthetic flatfields which are used as a calibrator of $T(\vec{r}, \lambda) \cdot QE(\vec{r}, \lambda)/g$.

As of today, no such dataset is available and the strategy has been to split the flatfielding into two steps: the pixel-to-pixel variation is corrected using a synthetic flatfield built from a set of broadband flatfields using a dome screen, while the QE calibration is performed using a single QE curve for the detector, obtained from collimated beam projector observations. 

We now combine two methods to determine the wavelength calibration for the grating.
A first, coarse, estimate of the dispersion relation is required to calibrate the imager's portion where the footprint of the spectrum lies. The pixel to wavelength transformation can be estimated by a visual inspection of known spectral features on the raw, uncalibrated spectrum, or predicted by knowing the geometry of the disperser. The grating spacing ($a$) and the distance ($d$) between the focal plane and the grating are related in the following way
\begin{equation}
i_{\mathrm{pixel}} = d \times \left(\frac{\lambda/a}{\sqrt{1-(\lambda/a)^2}}\right),
\end{equation}
where $i_{\mathrm{pixel}}$ is the pixel index.
The linear relation applied to determine a first order pixel to wavelength transformation is not a sufficiently accurate representation of the dispersion relation of the instrument.
A second-order solution of the pixel-to-wavelength transformation is performed by matching stellar and atmospheric features as expected from forward modeling to the observed spectrum. The residuals are below 1 nm. Second order light contamination appears in the spectrum above $\approx$\,740\,nm. A practical way of removing it is to place a high-pass filter in the light path, but this was not used for these data.

Source extraction is run on the trimmed, bias subtracted, flatfielded image.
Then an astrometric solution is determined and a match with a list of reference stars with known SEDs is performed. Stars with known SEDs can be obtained either from Space Telescope Imaging Spectrograph (STIS) CALSPEC data or from theoretical models. We used CALSPEC data in the following analyses. The SED is converted to flux units and downsampled to the resolution of the observations.
Extracting a spectrum then consists of two tasks: the determination of pixel to wavelength transformation and the flux estimation from the transverse profile of the spectrum. 

\subsection{The single image method}

Theoretically, a single image can be used to determine the atmospheric transmission at a given time and pointing, provided that two conditions are met:

\begin{enumerate}
\item There is a target in the field with a known SED,
\item The instrumental response of the telescope ($T_{tel}$) is calibrated.
\end{enumerate}

Then, the atmospheric transmission $T_{atmo}$ at a given wavelength is simply:
\begin{equation}
 T_{atmo}(\lambda) =  \frac{S_{obs}(\lambda) }{SED(\lambda) \times T_{tel}(\lambda)}, 
  \label{eq0}
\end{equation}
where $S_{obs}$ refers to the spectrum extracted from the image. This is a flux as a function of wavelength, determined either for one or both $m+1$ and $m-1$ orders. This depends on the location of the 0th order on the focal plane, and possibly on the disperser\footnote{A Ronchi grating has two first orders of interest; a blazed Ronchi only has one, but with more favorable flux ratio between direct light and first order.}. 

Given that this method requires that the telescope system response be known to a precision better than 1 \%, and given that our collimated beam projector device is still a prototype device, the current strategy to determine the atmospheric transmission is to combine observations of spectra in a time series with satellite data and a radiative transfer simulation.

\subsection{Time Series Method}

For this method, we determine atmospheric transparency by combining uncalibrated spectrophotometry with a radiative transfer simulation of the atmosphere (libradTran\footnote{\url{http://www.libradtran.org/doc/libRadtran.pdf}}) using input parameters from satellite data.
The satellite data are provided by the National Aeronautics and Space Administration (NASA) Modern-Era Retrospective analysis for Research and Applications, Version 2 (MERRA-2) project\footnote{\url{https://disc.sci.gsfc.nasa.gov/datasets?page=1&keywords=MERRA-2}}. A description of the variables can be found here\cite{BoLu2016}. From the literature\cite{StHi2007,BuCo2013}, it is found that atmospheric transparency and its temporal evolution can be decomposed into a small set of parameters:

\begin{itemize}
\item Molecular scattering optical depth,
\item Aerosol optical depth (AOD),
\item Ozone optical depth (O$_3$),
\item Precipitable Water Vapour (PWV).
\end{itemize}

\begin{figure*}[t]
\centering
\hspace*{-0.5cm}
 \includegraphics[width=3.25in]{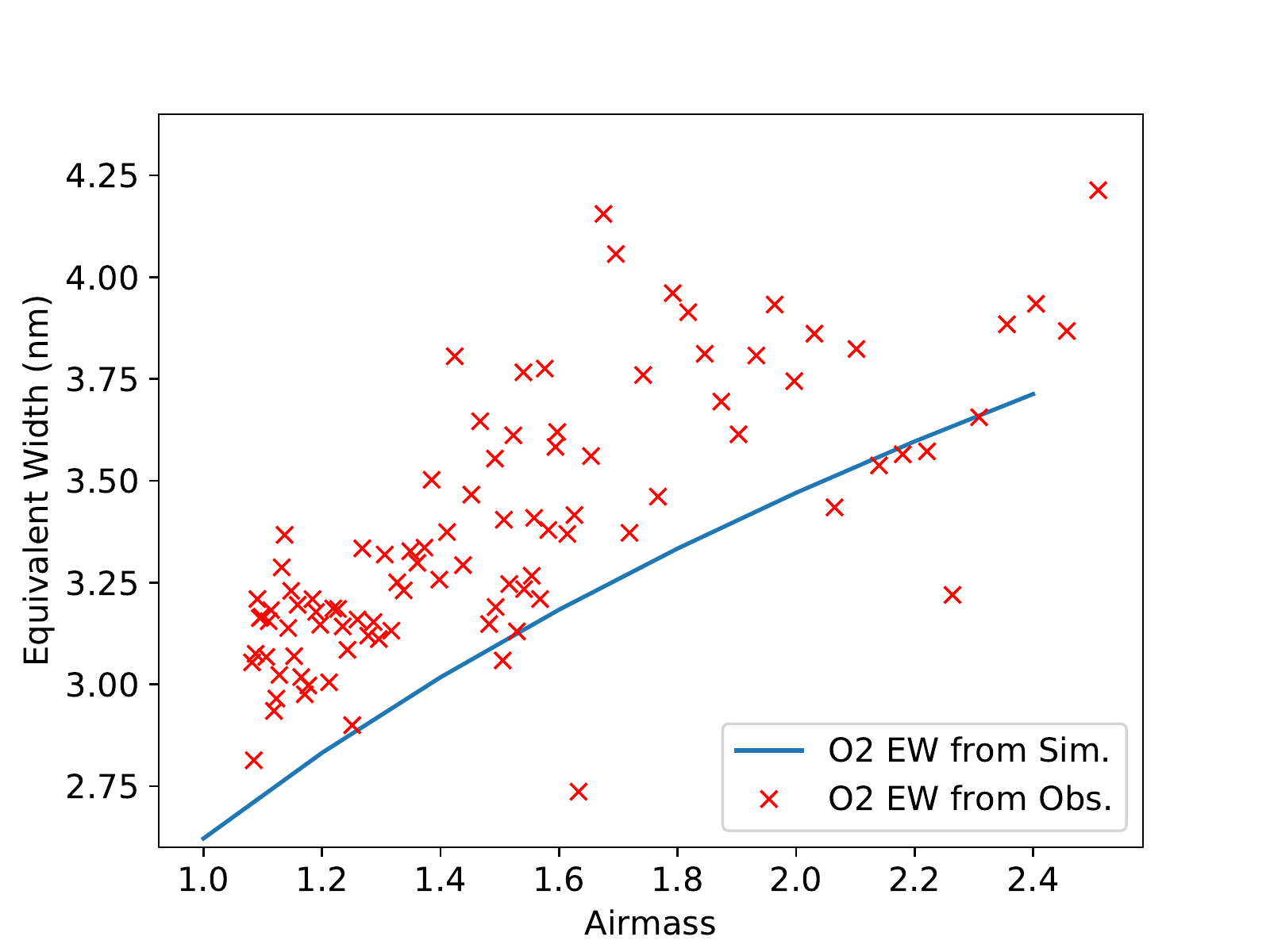}
 \includegraphics[width=3.25in]{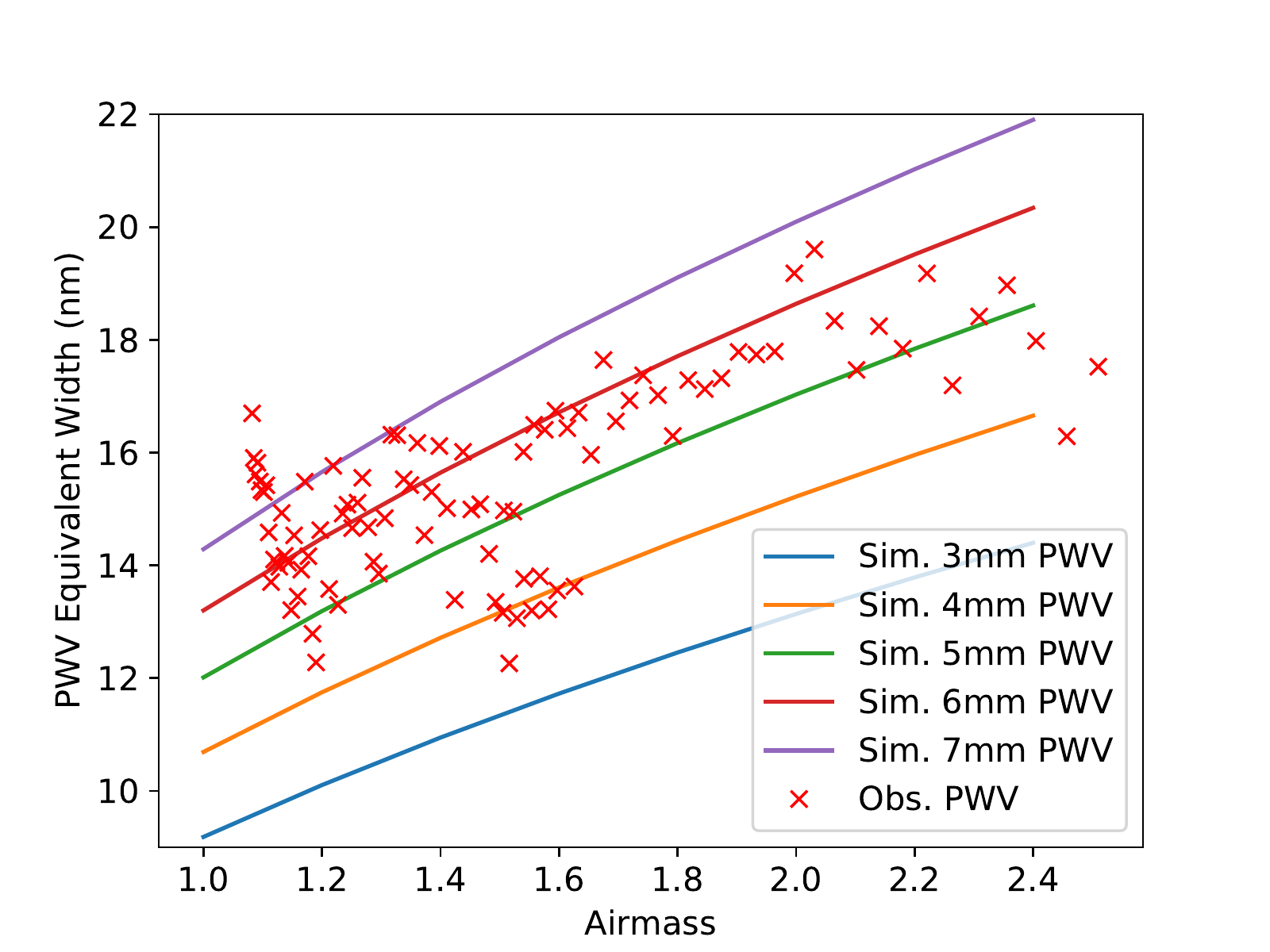}
 \caption{On the left is the measurement of O$_2$ equivalent width during the night of October 10, 2017 (red crosses) compared with the same equivalent width (EW) measurement applied to libRadTran simulations with airmasses covering the same range (solid line). On the right is determination of PWV during the night of October 10, 2017. The red crosses indicate measurements of EW in the range 880\,nm to 990\,nm. Colored lines indicates the same EW measurement applied to libRadTran simulations with airmasses covering the same range and with PWV ranging from 3\,mm to 7\,mm.}
 \label{fig:o2_pwv_ew}
\end{figure*}

The optical depth for scattering processes (Rayleigh and Mie scattering) is proportional to airmass. It also varies as a function of wavelength, and can be parametrized in the following way
\begin{equation}
\tau_R = \frac{X(h, \theta)}{ 2770\,\mathrm{g}/\mathrm{cm}^2} \left( \frac{400\,\mathrm{nm}}{\lambda} \right)^4
\end{equation}
where $X(h, \theta)$ is the air column depth as a function of height $h$ and scattering angle $\theta$.
The AOD is modeled by an empirical formula:
\begin{equation}
\tau_A = \beta \lambda^{(-\alpha)}
\end{equation}
where $\alpha$ is the wavelength exponent, and $\beta$ is the
turbidity coefficient. It is expected for $\alpha$ to vary between 0.5 and 1.5. From MERRA-2  data, $\beta$ is expected to vary between 0.02 and 0.12, with occasional large variations over a few hours time span. The ozone content impacts the atmospheric transparency curve in two distinct regions called the Huggins band (below 350 nm) and the Chappuis bands (500 nm to 700 nm).
The satellite data indicates that ozone concentrations at CTIO follow both seasonal and circadian variations, the distribution being between $6.5  \times  10^{18}$\,molecules cm$^{-2}$ (240 Dobson) and $8.6  \times  10^{18}$\,molecules cm$^{-2}$ (320 Dobson) most of the time. The total precipitable water vapor in a vertical column is measured in kg/m$^2$, or sometimes represented in mm. Its most significant impact is in the red, above 700\,nm. Its temporal evolution is correlated with the temperature and barometric pressure profiles in the vicinity of the telescope.

The analysis of the MERRA-2 data over the last few years allows us to identify several general trends :
\begin{itemize}
\item CTIO sits on a large east-west PWV gradient,
\item PWV and ozone follow circadian variations,
\item PWV and ozone are anti-correlated on an annual basis,
\item Large variations of AOD, PWV and O$_3$ can sometimes occur within a few hours timespan, 
\item O$_3$ and PWV gradients go along the same direction.
\end{itemize}
 
\begin{figure*}[t]
\centering
\hspace*{-0.5cm}
 \includegraphics[width=5in]{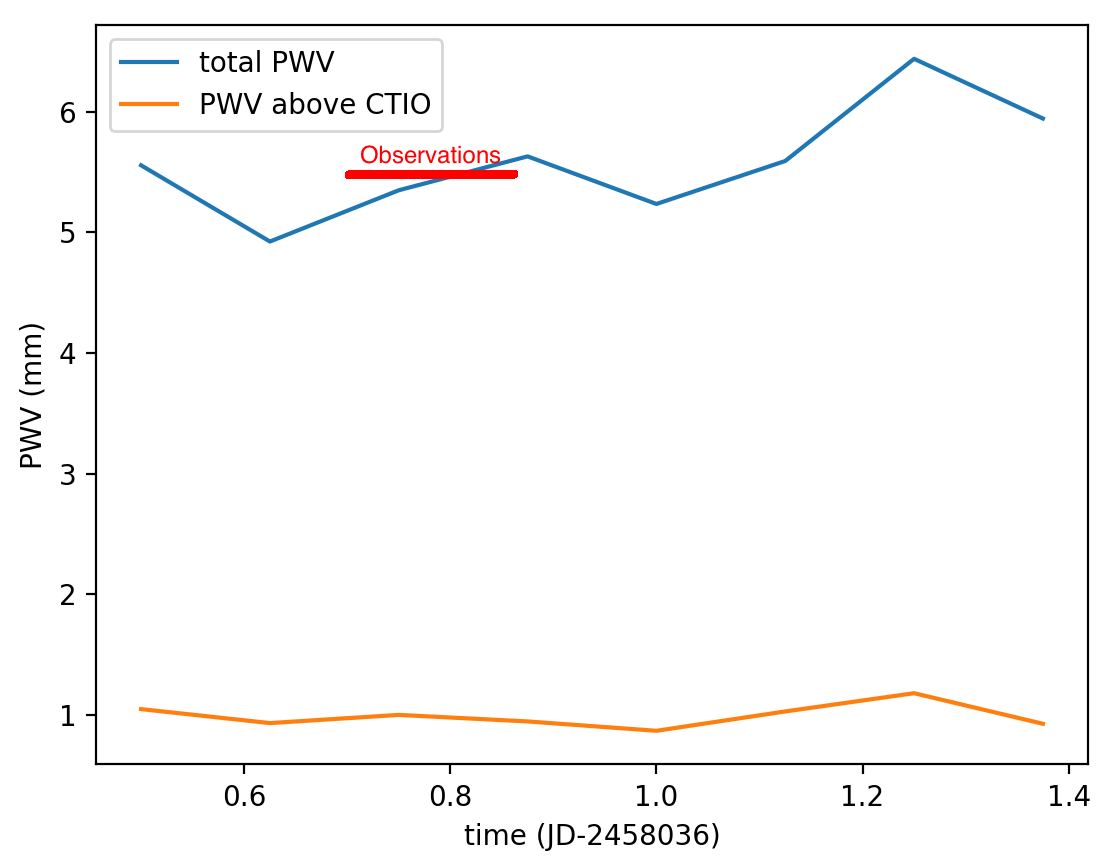}
 \caption{Comparison of the PWV measurement (large red band) with two different MERRA-2 tables: the blue curve corresponds to a longitude-latitude grid interpolation of the 2-D data at the CTIO location, while the orange curve is obtained from a 3-D interpolation: this is the latitude-longitude interpolation of the specific humidity, then integrated from CTIO altitude (2200\,m) up to the top layer of the atmosphere (JD = Julian Day).}
 \label{fig:pwv_ew2}
\end{figure*}

Our determination of an atmospheric transmission function curve is obtained by combining spectra, satellite data and simulation in the following. The ozone ($\textrm{O}_3$) is obtained from satellite data. Note that $\textrm{O}_2$ absorption is not a free parameter of the model. Instead the measurement of its equivalent width (EW) can be used as an estimator for biases in the method (see left of figure~\ref{fig:o2_pwv_ew}).
The precipitable water vapor (PWV) is determined by the measurement of an EW in the range 880\,nm to 990\,nm (red crosses, figure~\ref{fig:o2_pwv_ew}) from 86 spectra of Lamlep and 11 Ksi02Cet spectra observed during the night of October 10th, 2017. These values are then compared with the same EW measured on a libRadTran simulation with various PWV levels (right of figure~\ref{fig:o2_pwv_ew}, colored lines), from which we estimate the mean value for the night to be between 5 mm and 6 mm. The result can be compared with predictions inferred from satellite data, as shown in Figure~\ref{fig:pwv_ew2}. The blue curve corresponds to the total precipitable water vapor that is interpolated from a longitude-latitude 2-D grid. This is most likely an overestimation given the mountainous terrain of the CTIO site.  Some of the satellite pointings are taken over the Pacific ocean, thus integrating the PWV down to the sea level. To account for the altitude of the telescope site (2200 m), a second estimate is derived from a MERRA-2 3-D tables: Specific humidity is integrated over 59 bins from the telescope altitude, up to the top of the atmosphere. The result is indicated by the orange line on the same figure. These two results are to be taken as an indication of the range where the correct value should be expected.

\begin{figure*}[t]
\centering
\hspace*{-0.5cm}
 \includegraphics[width=5in]{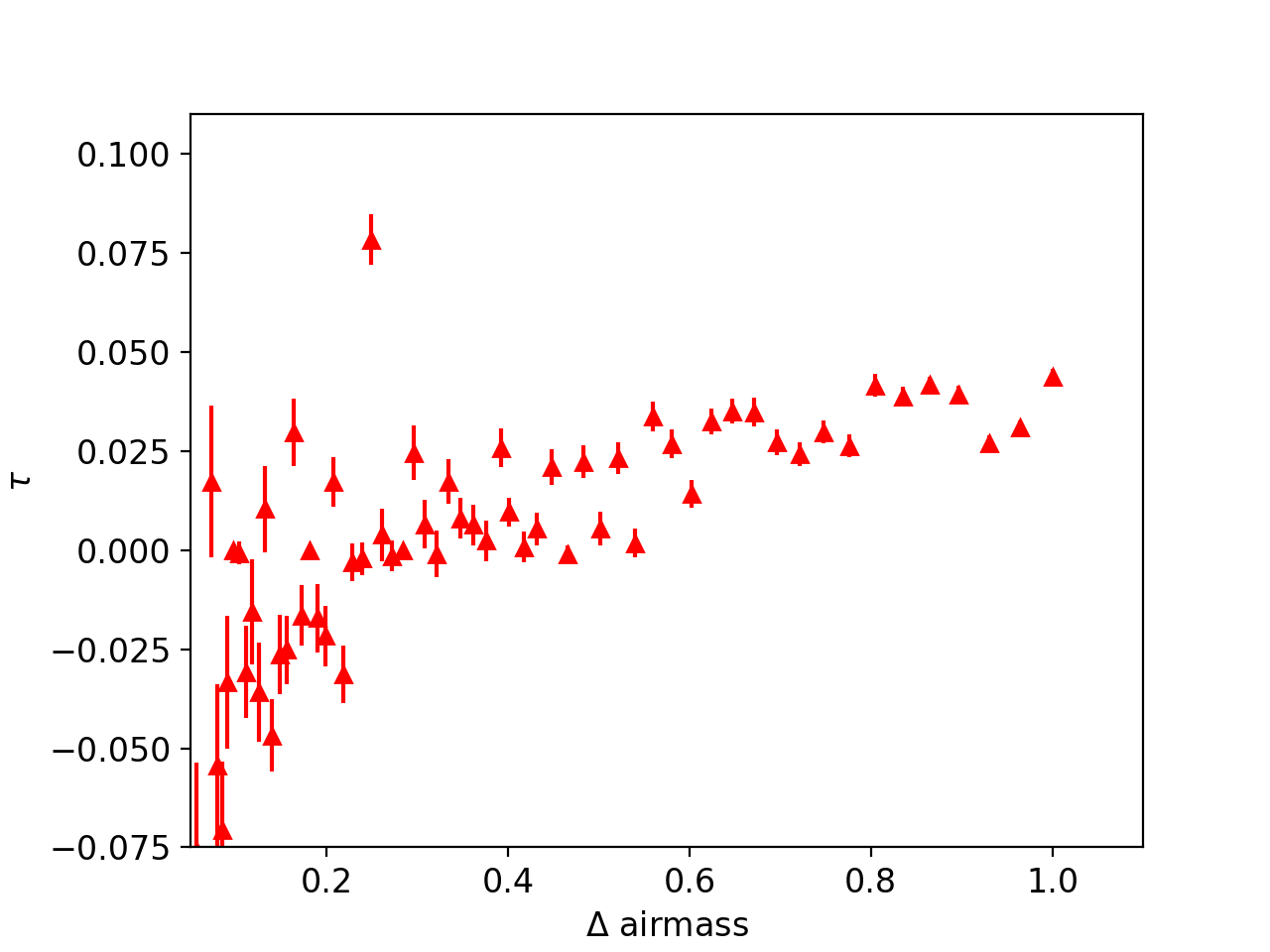}
 \caption{Aerosol optical depth $\tau$ for the night of October 10, 2017.}
 \label{fig:tau}
\end{figure*}

Lastly, we now determine the AOD by fitting its dependency as a function of airmass.
The spectrum $S$ of a standard star $SED$ is the product of
\begin{equation}
  S(\lambda, z, t) = SED(\lambda) \times T_{tel}(\lambda, t) \times T_{atmo}(\lambda, z, t) 
\end{equation}
By examining the same target at two different airmasses $z_1$, $z_2$ during a photometric night, we can relate $T_{atmo}^{z1}(\lambda)$ and $T_{atmo}^{z2}(\lambda)$
\begin{equation}
  \frac{S_{z1}(\lambda)}{S_{z2}(\lambda)}  =\frac{T_{atmo}^{z1}(\lambda)}{T_{atmo}^{z2}(\lambda)}.
  \label{obs}
\end{equation}
In addition, following equation 5 in Buton et al. \cite{BuCo2013},
\begin{equation}
T_{atmo}^{z}(\lambda) = 10^{-0.4 K_{atmo}(\lambda, z)}.
\end{equation}
In a region free of telluric lines,
\begin{equation}
K_{atmo}(\lambda, z) = z k_r + z k_A + z k_{O_3}
\end{equation}
where $k_r$ is the Rayleigh scattering, $k_A$ the aerosol scattering and $k_{O_3}$ is the ozone absorption. Assuming aerosols follow an inverse power law with wavelength:
\begin{equation}
k_{A}(\lambda) = \tau \lambda^{-\alpha}
\end{equation}
Equation \ref{obs} can then be rewritten as
\begin{equation}
  \frac{S_{z_1}(\lambda)}{S_{z_2}(\lambda)}  =\frac{10^{-0.4 z_1 (k_r(\lambda) + k_{o3}(\lambda))} \cdot 10^{-0.4 z_1 \tau \lambda^{-\alpha}} }{10^{-0.4 z_2 (k_r(\lambda) + k_{o3}(\lambda))} \cdot 10^{-0.4 z_2 \tau \lambda^{-\alpha}}}.  
\end{equation}
Using a radiative transfer simulation (without aerosols) of the two observations at two airmasses:
\begin{equation}
  \frac{\Big(S_{z_1}(\lambda)/(S_{z_2}(\lambda)\Big)}{\Big({T_{atmo}^{z_1}}_{sim}^{noA}(\lambda)/({T_{atmo}^{z_2}}_{sim}^{noA}(\lambda)\Big)}  = 10^{-0.4 (z_2 - z_1)\tau \lambda^{-\alpha}}
\end{equation}
where $\tau$ is the AOD, and $\alpha$, the Angstrom exponent are the two parameters to be adjusted on spectra in a time series. The AOD measurement for the night of October 10, 2017 is shown in figure \ref{fig:tau}. This is using a set of observation of Lamlep as a stellar light source to back-illuminate the atmosphere at various airmasses. The result is in a range between 0.00 and 0.05, which is slightly smaller than the MERRA-2 values $\tau_{550nm}$ in the range 0.05-0.09.

This exercise is a first attempt to demonstrate the ability of slitless, low-resolution spectrometry, to monitor atmospheric parameters on a nightly basis.
The results that are obtained here, using the CTIO 0.9\,m telescope, are found to be in agreement with what can be interpolated from satellite observations. The precision is currently of the same order of magnitude of the variations that are expected during photometric nights, and a systematic comparison with a synthetic photometry based on real time atmospheric curves shall benefit from a dedicated telescope, such as the LSST auxiliary telescope.

\section{Conclusion}
\label{sec:conclusion}

In this paper, we have discussed progress towards the measurement of a telescope's instrumental response function versus wavelength and atmospheric transmission.
Using a collimated beam projector, we showed consistency with previous measurements of broadband filters at the telescope. We also described a number of outstanding technical issues that will need to be addressed in the future. In addition, we described how CALSPEC stars can be used as calibrated light sources to illuminate the atmosphere and measure its transmission. We described how a combination of Ronchi grating observations with atmospheric modeling codes can be used to derive atmospheric properties as a function of time.

This measurement will be informative for plans for LSST calibration. The configuration of the SMARTS 0.9\,m telescope is similar to the instrument and site conditions of the LSST auxiliary telescope (AuxTel) currently under construction on Cerro Pach\'{o}n, which neighbors Cerro Tololo.  The LSST AuxTel will monitor spatial and temporal variability of atmospheric transmission in real time. The strategy shall be to use this telescope to acquire spectra from standard stars near the science field and use them to correct the effective passband model, accounting for the short time scale variations of atmospheric transmission. In addition, a larger collimated beam projector is under construction to perform dedicated throughput measurements for LSST. In the end, as the goal is to provide the instrument throughput as a function of source location, this will require rastering our multiplexed source around the focal plane.
In addition, absolute calibration will require rastering the collimated beam projector footprint around the primary (appropriately overlapping) so as to scan the entire collecting area of LSST.

\acknowledgments
MC was supported by the David and Ellen Lee Postdoctoral Fellowship at the California Institute of Technology.
 CWS is grateful to the DOE Office 
of Science for their support under award DE-SC0007881.
NM is supported by National Science Foundation Graduate Research Fellowship Program, under NSF grant number DGE 1745303.  NM thanks the LSSTC Data Science Fellowship Program, his time as a Fellow has benefited this work.

$^*$Note: References are made to certain commercially available products in this paper to adequately specify the experimental
procedures involved. Such identification does not imply recommendation or endorsement by the National Institute of Standards and
Technology, nor does it imply that these products are the best for the purpose specified.
\bibliography{references} 

\begin{thebibliography}{10}

\bibitem{BoGo2014}
Bohlin, R.~C., Gordon, K.~D., and Tremblay, P.-E., ``Techniques and review of
  absolute flux calibration from the ultraviolet to the mid-infrared,'' {\em
  Publications of the Astronomical Society of the Pacific}~{\bf 126}(942),  711
  (2014).

\bibitem{StDo2010}
{Stubbs}, C.~W., {Doherty}, P., {Cramer}, C., {Narayan}, G., {Brown}, Y.~J.,
  {Lykke}, K.~R., {Woodward}, J.~T., and {Tonry}, J.~L., ``{Precise Throughput
  Determination of the PanSTARRS Telescope and the Gigapixel Imager Using a
  Calibrated Silicon Photodiode and a Tunable Laser: Initial Results},'' {\em
  The Astrophysical Journal Supplements}~{\bf 191},  376--388 (Dec. 2010).

\bibitem{AlBu2016}
Allende~Prieto, C. and del Burgo, C., ``New bright optical spectrophotometric
  standards: A-type stars from the stis next generation spectral library,''
  {\em Monthly Notices of the Royal Astronomical Society}~{\bf 455}(4),
  3864--3870 (2016).

\bibitem{Boh2016}
Bohlin, R.~C., ``Perfecting the photometric calibration of the acs ccd
  cameras,'' {\em The Astronomical Journal}~{\bf 152}(3),  60 (2016).

\bibitem{NaAx2016}
Narayan, G., Axelrod, T., Holberg, J.~B., Matheson, T., Saha, A., Olszewski,
  E., Claver, J., Stubbs, C.~W., Bohlin, R.~C., Deustua, S., and Rest, A.,
  ``Toward a network of faint da white dwarfs as high-precision
  spectrophotometric standards,'' {\em The Astrophysical Journal}~{\bf 822}(2),
   67 (2016).

\bibitem{KrSu2017}
Krisciunas, K., Suntzeff, N.~B., Kelarek, B., Bonar, K., and Stenzel, J.,
  ``Spectrophotometry of very bright stars in the southern sky,'' {\em
  Publications of the Astronomical Society of the Pacific}~{\bf 129}(975),
  054504 (2017).

\bibitem{StTo2006}
Stubbs, C.~W. and Tonry, J.~L., ``Toward 1\% photometry: End-to-end calibration
  of astronomical telescopes and detectors,'' {\em The Astrophysical
  Journal}~{\bf 646}(2),  1436 (2006).

\bibitem{BoDi2001}
{Bohlin}, R.~C., {Dickinson}, M.~E., and {Calzetti}, D., ``{Spectrophotometric
  Standards from the Far-Ultraviolet to the Near-Infrared: STIS and NICMOS
  Fluxes},'' {\em The Astronomical Journal}~{\bf 122},  2118--2128 (Oct. 2001).

\bibitem{BoMe2017}
{Bohlin}, R.~C., {M{\'e}sz{\'a}ros}, S., {Fleming}, S.~W., {Gordon}, K.~D.,
  {Koekemoer}, A.~M., and {Kov{\'a}cs}, J., ``{A New Stellar Atmosphere Grid
  and Comparisons with HST/STIS CALSPEC Flux Distributions},'' {\em The
  Astronomical Journal}~{\bf 153},  234 (May 2017).

\bibitem{StHi2007}
{Stubbs}, C.~W., {High}, F.~W., {George}, M.~R., {DeRose}, K.~L., {Blondin},
  S., {Tonry}, J.~L., {Chambers}, K.~C., {Granett}, B.~R., {Burke}, D.~L., and
  {Smith}, R.~C., ``{Toward More Precise Survey Photometry for PanSTARRS and
  LSST: Measuring Directly the Optical Transmission Spectrum of the
  Atmosphere},'' {\em Publications of the Astronomical Society of the
  Pacific}~{\bf 119},  1163--1178 (Oct. 2007).

\bibitem{StTo2012}
{Stubbs}, C.~W. and {Tonry}, J.~L., ``{Addressing the Photometric Calibration
  Challenge: Explicit Determination of the Instrumental Response and
  Atmospheric Response Functions, and Tying it All Together},'' {\em ArXiv
  e-prints}  (June 2012).

\bibitem{Houghton1977}
Houghton, J., ``{The Physics of Atmospheres},'' (1977).

\bibitem{Slater1980}
Slater, P., ``Remote sensing, optics and optical systems,'' (1980).

\bibitem{CoAb2016}
Coughlin, M., Abbott, T. M.~C., Brannon, K., Claver, C., Doherty, P.,
  Fisher-Levine, M., Ingraham, P., Lupton, R., Mondrik, N., and Stubbs, C., ``A
  collimated beam projector for precise telescope calibration,'' {\em Proc.
  SPIE}~{\bf 9910},  99100V--99100V--10 (2016).

\bibitem{DoTa2010}
Doi, M., Tanaka, M., Fukugita, M., Gunn, J.~E., Yasuda, N., Željko Ivezić,
  Brinkmann, J., de~Haars, E., Kleinman, S.~J., Krzesinski, J., and Leger,
  R.~F., ``Photometric response functions of the sloan digital sky survey
  imager,'' {\em The Astronomical Journal}~{\bf 139}(4),  1628 (2010).

\bibitem{ToSt2012}
{Tonry}, J.~L., {Stubbs}, C.~W., {Lykke}, K.~R., {Doherty}, P., {Shivvers},
  I.~S., {Burgett}, W.~S., {Chambers}, K.~C., {Hodapp}, K.~W., {Kaiser}, N.,
  {Kudritzki}, R.-P., {Magnier}, E.~A., {Morgan}, J.~S., {Price}, P.~A., and
  {Wainscoat}, R.~J., ``{The Pan-STARRS1 Photometric System},'' {\em The
  Astrophysical Journal}~{\bf 750},  99 (May 2012).

\bibitem{Dok2001}
van Dokkum, P.~G., ``Cosmic‐ray rejection by laplacian edge detection,'' {\em
  Publications of the Astronomical Society of the Pacific}~{\bf 113}(789),
  1420 (2001).

\bibitem{BoLu2016}
Bosilovich, M.~G., Lucchesi, R., and Suarez, M., ``{MERRA-2: File
  Specification},'' {\em GMAO Office Note No. 9 (Version 1.1)} ,  73 (2016).

\bibitem{BuCo2013}
{Buton}, C., {Copin}, Y., {Aldering}, G., {Antilogus}, P., {Aragon}, C.,
  {Bailey}, S., {Baltay}, C., {Bongard}, S., {Canto}, A., {Cellier-Holzem}, F.,
  {Childress}, M., {Chotard}, N., {Fakhouri}, H.~K., {Gangler}, E., {Guy}, J.,
  {Hsiao}, E.~Y., {Kerschhaggl}, M., {Kowalski}, M., {Loken}, S., {Nugent}, P.,
  {Paech}, K., {Pain}, R., {P{\'e}contal}, E., {Pereira}, R., {Perlmutter}, S.,
  {Rabinowitz}, D., {Rigault}, M., {Runge}, K., {Scalzo}, R., {Smadja}, G.,
  {Tao}, C., {Thomas}, R.~C., {Weaver}, B.~A., {Wu}, C., and {Nearby SuperNova
  Factory}, ``{Atmospheric extinction properties above Mauna Kea from the
  Nearby SuperNova Factory spectro-photometric data set},'' {\em Astronomy and
  Astrophysics}~{\bf 549},  A8 (Jan. 2013).

\end{thebibliography}
\bibliographystyle{spiebib} 
 
\end{document}